\documentclass[a4paper,11pt]{article}
\usepackage{jinstpub} 
\usepackage{lineno}
\usepackage{comment}
\usepackage[normalem]{ulem}

\title{\boldmath Progress in Muon Cooling and Moderation Techniques}

\author[a]{C.~Zhang}
\author[b]{K.~S.~Khaw}
\author[c]{C.~Rogers}

\affiliation[a]{University of Liverpool,\\
Liverpool, United Kingdom}
\affiliation[b]{State Key Laboratory of Dark Matter Physics, Key Laboratory for Particle Astrophysics and Cosmology (MOE), Shanghai Key Laboratory for Particle Physics and Cosmology (SKLPPC),  Tsung-Dao Lee Institute \& School of Physics and Astronomy, Shanghai Jiao Tong University, Shanghai, China}
\affiliation[c]{Rutherford Appleton Laboratory,\\
Didcot OX11 0QX, United Kingdom}

\emailAdd{ce.zhang@liverpool.ac.uk, kimsiang84@sjtu.edu.cn, chris.rogers@stfc.ac.uk}

\abstract{

Muon beams are essential tools for a wide range of applications such as the search for rare muon decays and muon spectroscopy. However, conventional muon beams produced from pion decay have large emittance and broad momentum spread. Conventional beam-cooling techniques cannot improve the beam quality significantly owing to the 2.2~\textmu s muon mean lifetime. This has motivated the development of dedicated muon beam cooling and moderation methods. This review surveys the principal approaches, including ionisation cooling for high-brightness muon beams, low-energy muon moderation and frictional cooling for slow muon production, and laser ionisation of thermal muonium for ultra-low-emittance muon sources. The challenges associated with producing slow negative muons and emerging concepts based on cyclotron trapping and muon-catalysed fusion are also discussed. These complementary techniques offer promising pathways towards next-generation muon facilities and precision experiments.

}


\begin{document}
\maketitle
\flushbottom

\section{Introduction}
\label{sec:intro}

Muon beams have become powerful tools, enabling precision experiments such as measurement of the muon anomalous magnetic moment (muon $g\!-\!2$)~\cite{Muong-2:2006rrc,Muong-2:2015xgu,Muong-2:2025xyk,Hertzog:2025ssc,abe_new_2019}, future high-energy muon colliders~\cite{Neuffer:1994sc,Palmer:2014nza,Neuffer:2018yof,Boscolo:2018tlu},  muon spin rotation spectroscopy ($\mu$SR)~\cite{Blundell:1999zz,Reotier_1997,hillier2022muon}, muonic atom spectroscopy~\cite{Pohl:2010zza,Antognini:2013txn}, and other studies in materials and life sciences~\cite{Bonechi:2019ckl}.
The scientific reach of these applications critically depends on the available muon
beam quality, including its intensity, emittance, momentum spread, polarisation, and energy.

In the context of muon beams, the terms cooling and moderation are used to describe complementary classes of beam manipulation. The former refers to schemes that reduce the beam phase-space volume, or equivalently increase the phase-space density and beam brightness, whereas the latter denotes processes that primarily lower the muon kinetic energy to enable subsequent capture, transport, or application. The distinction is not always sharp, and several techniques discussed in this review exhibit characteristics of both cooling and moderation.

Because muon beams are usually produced as tertiary beams through the decay of secondary pions, they are inherently characterised by large emittance, broad momentum spread, and relatively low phase-space density. Consequently, beam manipulation techniques that improve beam quality are important for their intended applications. Unlike stable charged particles, however, muons have a lifetime of only $\tau_\mu = 2.2~\mu$s at rest. Conventional beam-cooling methods, such as electron cooling~\cite{Parkhomchuk:2000xt} or stochastic cooling~\cite{Mohl:1980jb,Marriner:2003mn}, are generally too slow to be effective on this timescale, motivating the development of dedicated cooling and moderation techniques.

Muon beam cooling and moderation has gained interest following developments in muon beam production itself~\cite{Budker:1978gp,Skrinsky:1981ht}. Muons arising from decay of pions in flight yield higher energy beams while decay of pions near the surface of production targets yield lower energy and highly polarised muon beams. Beam channels select the desired momentum, suppress contamination and preserve polarisation. These developments established a practical framework for muon science, but also made clear that conventional beam transport was not sufficient for experiments requiring colder or more highly compressed beams.

A sequence of slow-muon concepts was proposed over the following decades~\cite{Nagamine:2003sv}. 
Early ideas included thermal muonium ionisation in hot tungsten targets, cold-moderator schemes based on frozen noble-gases, frictional cooling in the low-energy region where the stopping power rises as the muon energy decreases, and beam-cooling or phase-space-compression concepts using electromagnetic confinement and applied electric fields. Beyond these low-energy approaches, more ambitious schemes such as inverse cyclotron cooling, PRISM-type phase-space rotation, and ionisation cooling were developed to address higher-energy muon beams. Many of these concepts laid the conceptual and technical foundations for later advances in the field, and their development has evolved in parallel with improvements in primary proton accelerators and the corresponding increase in available muon intensity~\cite{Hotchi:2021ztk}. Current active facilities include Japan Proton Accelerator Research Complex (J-PARC), Paul Scherrer Institute (PSI), ISIS, Fermilab, TRIUMF, CERN, and Jefferson Lab~\cite{Strasser:2014dsa,kanda2023ultra,Berg:2015wna,Marshall:1992,Maso:2023zjp,Hillier:2019szm,MICE:2012sli,MICE:2013qmn,Ganguly:2022ufq,Doble:1994np,Achenbach:2025ynn}.

Depending on the intended application, different beam manipulations are required, and no single technique is optimal for all muon sources or experiments. 
In this review, we focus on the several approaches which have emerged as the most prominent and application-driven: ionisation cooling (Section~\ref{sec:ionisation}), developed primarily in the context of muon colliders and neutrino sources; slow muon moderation and frictional cooling (Section~\ref{sec:friction}), aimed at producing low-energy muon beams; and laser ionisation cooling (Section~\ref{sec:laser}), originally proposed for muon $g\!-\!2$ measurement at J-PARC~\cite{Otani:2015lra}. These methods will be discussed in separate sections of this review. Some techniques exhibit features of both cooling and moderation, since they combine energy reduction with a degree of phase-space manipulation.

\begin{figure}[htbp]
    \centering
    \includegraphics[width=0.9\linewidth]{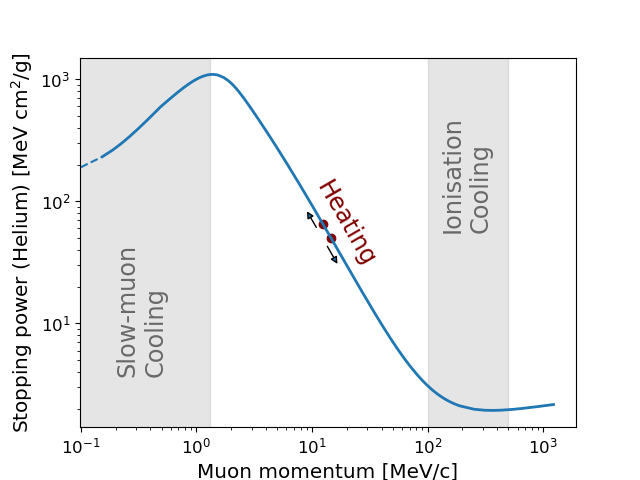}
    \caption{Positive muon stopping power as a function of momentum. The curve is obtained from NIST PSTAR proton stopping-power data, mapped to muons at the same $\beta\gamma$. It agrees with PDG muon data for kinetic energies above $1~\mathrm{MeV}$; below this energy, it represents an estimate based on proton data. Negative muons are vulnerable to nuclear capture at momenta below a few MeV/c (not shown).}
    \label{fig:dedx}
\end{figure}

The disparity between slow-muon and fast-muon techniques arises because of the momentum dependence of the energy loss as shown in Fig.~\ref{fig:dedx}. At low energy positive muons passing through matter tend to experience energy compression; muons that have higher energy are stopped more quickly. At intermediate energy, but below the minimum ionising energy, muons experience a dramatic increase in energy spread; muons that have a lower energy are stopped more quickly, leading to a growing energy spread that is hard to counteract. 
\textcolor{black}{Near the minimum-ionizing region, ionization cooling can effectively reduce the beam’s transverse emittance. At energies well above this region, stochastic energy-loss fluctuations can become significant relative to the cooling effect.}

It is also important to note that most of the techniques mentioned above were developed primarily for positive muons, although ionisation cooling can also be applied to negative muons. Negative muons behave differently because nuclear capture introduces an additional loss channel and the stopping power needs to be corrected (the Barkas correction), thereby altering both the practical limits and the optimal strategy for cooling and moderation. Dedicated schemes for negative muons, such as using the muon-catalyzed fusion, will therefore be discussed separately in Section~\ref{sec:negative}.

\section{Ionisation Cooling}
\label{sec:ionisation}

\subsection{Principle}

Ionisation cooling is the only high energy muon cooling technique that has been experimentally demonstrated and remains the baseline technology for future muon colliders and for neutrino factories~\cite{ISSPhysicsWorkingGroup:2007lul}. Unlike frictional cooling or laser-ionisation cooling, which are primarily intended for low-energy muons, ionisation cooling operates on relativistic muon beams with typical momenta of a few hundred~MeV/c. Its primary objective is to reduce the transverse emittance and, through emittance exchange, ultimately achieve six-dimensional (6D) phase-space compression within the short muon lifetime.

\begin{figure}[htbp]
    \centering
    \includegraphics[width=0.9\linewidth]{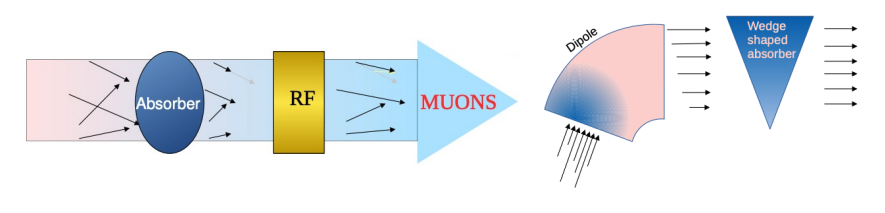}
    \caption{Principle of ionisation cooling. Muons lose momentum in a low-$Z$ absorber and are subsequently re-accelerated in RF cavities, restoring only the longitudinal momentum and thereby reducing the transverse emittance. Wedge absorbers placed in dispersive regions enable longitudinal--transverse emittance exchange for 6D cooling.}
    \label{fig:ionisation}
\end{figure}

The initial idea of ionisation cooling was illustrated in Fig.~\ref{fig:ionisation}. As muons traverse a low-$Z$ absorber, such as liquid hydrogen or lithium hydride, they lose momentum through ionisation energy loss~\cite{Neuffer:1983jr}. Since the energy loss reduces both the longitudinal and transverse momentum components, radio-frequency (RF) cavities are placed downstream of the absorber to restore only the longitudinal momentum. Repeating this sequence over many cooling cells progressively reduces the transverse emittance while maintaining the average beam energy.

The evolution of the normalised transverse emittance is governed by the competition between cooling due to ionisation energy loss and heating arising from multiple Coulomb scattering~\cite{Neuffer:1983jr},

\begin{equation}
\frac{d\varepsilon_N}{ds}
=
-\frac{\varepsilon_N}{\beta^2 E_\mu}
\left\langle\frac{dE}{ds}\right\rangle
+
\frac{\beta_\perp (13.6~\mathrm{MeV})^2}
{2 \beta^3 E_\mu m_\mu X_0},
\label{eq:ionization_cooling}
\end{equation}
where $\varepsilon_N$ is the normalised transverse emittance, $s$ is the path length, $\beta=v/c$, $E_\mu$ is the muon total energy, $\langle dE/ds\rangle$ is the mean ionisation energy loss, $\beta_\perp$ is the betatron function at the absorber, $m_\mu$ is the muon mass, and $X_0$ is the radiation length of the absorber material. The first term represents emittance reduction through ionisation energy loss, whereas the second term describes emittance growth caused by multiple Coulomb scattering.

As cooling proceeds, the two competing processes eventually balance each other, leading to an equilibrium emittance. By setting $d\varepsilon_N/ds=0$, Eq.~(\ref{eq:ionization_cooling}) yields

\begin{equation}
\varepsilon_{N,\mathrm{eq}}
=
\frac{\beta_\perp (13.6~\mathrm{MeV})^2}
{2 \beta m_\mu X_0
\left\langle dE/ds \right\rangle},
\label{eq:equilibrium_emittance}
\end{equation}
which represents the minimum achievable normalised transverse emittance for a given lattice configuration and absorber material. Equation~(\ref{eq:equilibrium_emittance}) shows that stronger transverse focusing (smaller $\beta_\perp$) and absorber materials with longer radiation lengths can significantly improve the cooling performance, motivating the use of low-$Z$ materials such as liquid hydrogen and lithium hydride.

Ionisation cooling primarily reduces the transverse emittance. To achieve longitudinal cooling and full 6D phase-space compression, wedge absorbers are introduced in regions with finite dispersion. Higher-momentum muons traverse a greater absorber thickness and therefore lose more energy than lower-momentum particles, enabling emittance exchange between the longitudinal and transverse degrees of freedom. This principle forms the basis of 6D cooling channels proposed for future muon colliders.

Small transverse emittance may be achieved by focusing to very small $\beta_\perp$ using particularly strong solenoids operating on low momentum muon beams, which focus more readily. Muons are passed through an absorber embedded within the solenoid to achieve cooling. This is proposed as a final cooling system before muons are accelerated to collision in the muon collider. This cooling system operates in the heating region shown in Fig. \ref{fig:dedx}. The longitudinal emittance of the beam grows rapidly while the transverse emittance decreases. Overally luminosity is improved.


\subsection{Cooling Channel Designs}

The first experimental demonstration of ionisation cooling was carried out by the Muon Ionisation Cooling Experiment (MICE) at the Rutherford Appleton Laboratory~\cite{MICE:2019jkl,MICE:2023vpa}. Rather than constructing a complete cooling channel, MICE focused on demonstrating the cooling performance of a section of a cooling cell. Individual muons were tracked upstream and downstream of the absorber using high-precision spectrometers, enabling direct measurements of their phase-space coordinates before and after the absorber. Measurements with lithium hydride and liquid hydrogen absorbers showed a clear increase in beam core density together with a reduction in transverse emittance, providing the first experimental confirmation that ionisation cooling works as predicted. Since no RF re-acceleration was included, MICE demonstrated transverse cooling only, but nevertheless established the experimental foundation for future ionisation-cooling facilities.

Building upon the success of MICE, current research has shifted towards realizing practical cooling channels suitable for collider applications~\cite{Rogers:2023nqc}. The next major milestone is the construction of a cooling demonstrator incorporating repeated cooling cells, RF re-acceleration, and simultaneous transverse and longitudinal cooling~\cite{Jurj:2024dnm,Schulte:2025ity,Jurj:2025pra}. Compared with MICE, such a demonstrator would operate with bunched beams and multiple cooling stages, providing the first integrated demonstration of sustained 6D ionisation cooling. Recent rectilinear cooling-channel designs~\cite{Zhu:2024vfe} have shown approximately a factor-of-two reduction in the 6D emittance while maintaining a transmission of about 90\%, marking an important step from proof-of-principle experiments towards realistic collider cooling systems. Recent studies have addressed engineering aspects of these channels, including RF cavity configurations, magnet design, and tolerance analyses.

Looking further ahead, ionisation cooling forms an integral part of the baseline muon collider production chain. Although significant engineering challenges remain, particularly in high-gradient RF cavities operating in strong magnetic fields and high-field superconducting magnets, ionisation cooling remains the most mature and scalable solution for producing the high-brightness muon beams required by future muon colliders.

\section{Slow-Muon Moderation and Cooling}
\label{sec:friction}

\subsection{Low-energy muon moderation}

The production of low-energy ($\sim$eV) muon beams relies on the moderation of conventional MeV surface muons. Unlike beam degradation, which merely reduces particle energy and yields a broad continuous spectrum, a moderator exploits the energy-dependent stopping power of selected materials to generate a narrow distribution of epithermal muons with kinetic energies around $15~\mathrm{eV}$. Such slow-muon beams form the basis of low-energy $\mu$SR (LE-$\mu$SR), enabling depth-resolved studies of thin films, multilayers, and near-surface regions that are inaccessible with conventional surface muon beams~\cite{niedermayer1999direct}.

\textcolor{black}{The moderateration of positive muons was first demonstrated at TRIUMF in 1986, where a slow component with an energy of approximately 10\,eV was observed emerging from thin solid moderators of lithium fluoride and quartz~\cite{Harshman:1986zz}. In 1987, the production of slow positive muons was demonstrated using solid rare-gas moderators, including neon, argon, krypton, and xenon~\cite{Harshman:1987zz,Morris:1990}.} This technique is implemented today at the LEM (Low Energy Muon) beamline, the world's only facility dedicated to producing intense epithermal $\mu^+$ beams for condensed-matter research~\cite{Morenzoni:1994,Prokscha:2008zz}. The most widely used moderators there consist of thin ($\sim100$--$500~\mathrm{nm}$) layers of solid noble gases (Ne, Ar, Kr) or solid nitrogen deposited onto a cryogenic substrate, typically maintained below $20~\mathrm{K}$~\cite{Prokscha:2008zz}. Surface muons ($28~\mathrm{MeV}$/c) are first degraded in a backing foil to energies of several tens of keV before entering the moderator layer, where they lose energy through ionisation and charge-exchange processes as they propagate through the solid.

At kinetic energies below a few keV, charge-exchange interactions become increasingly important, driving repeated transitions between $\mu^+$ and muonium. As the muon energy approaches a few tens of eV, the electronic stopping power drops sharply: because solid noble gases form weakly bound van der Waals solids with very limited low-energy excitation channels, further inelastic energy loss is strongly suppressed. Muons remaining in the $\mu^+$ state can therefore diffuse over distances of up to $\sim100~\mathrm{nm}$ and escape the moderator with only minimal additional energy loss, producing a characteristic emission peak centered around $15~\mathrm{eV}$ with an energy spread of approximately $20~\mathrm{eV}$ (FWHM)~\cite{Morenzoni:2004}.

The moderation efficiency is intrinsically limited by several competing processes. Only a small fraction ($10^{-5}$--$10^{-4}$) of incident surface muons emerge as slow muons, the exact value depending on the moderator material and the incident beam momentum spread~\cite{Prokscha:2008zz}. Charge-exchange reactions during slowing-down further deplete the $\mu^+$ fraction by converting muons into muonium, and the majority of the incident beam exits the moderator as degraded muons with energies of several hundred keV. At LEM, an electrostatic transport and focusing system separates the epithermal $\mu^+$ component from this fast-muon background and refocuses it onto the experimental sample, with the implantation energy tunable from a few hundred eV to several tens of keV --- corresponding to implantation depths from a few nm to several hundred nm in condensed matter. At the sample position, LEM currently delivers a rate of order $10^3~\mu^+/\mathrm{s}$ (up to $\sim4.5\times10^3~\mu^+/\mathrm{s}$)~\cite{Prokscha:2008zz,Khaw:2015eya}, against the $\sim10^8~\mu^+/\mathrm{s}$ surface muon flux entering the beamline.

Crucially, the moderation process preserves the intrinsic high spin polarisation of the surface muon beam, so the resulting low-energy $\mu^+$ beam remains highly polarised despite the drastic reduction in kinetic energy. This is a feature that underlies LEM's utility for depth-resolved studies of magnetism, superconductivity, and interface phenomena in thin-film heterostructures.

\subsection{Frictional Cooling}

Frictional cooling~\cite{Muhlbauer:1999sc} represents a different approach from conventional ionisation cooling, aiming to reduce the muon kinetic energy directly into the sub-keV regime while simultaneously compressing the phase-space volume. In contrast to ionisation cooling, which relies on the balance between ionisation energy loss and multiple Coulomb scattering at relativistic energies, frictional cooling exploits the velocity dependence of the stopping power at muon momentum $\lesssim$ 1 MeV/c, where $\mathrm{d}E/\mathrm{d}x$ \emph{increases} with increasing energy: by applying an electric field that compensates the average energy loss in matter, muons can be driven towards a stable low-energy equilibrium. 

The muCool scheme pursued at PSI pushes this concept one step further: rather than maintaining muons at a keV-scale equilibrium energy, the muons are stopped completely in a helium gas target, and the resulting thermalised muon ``swarm'' is then compressed in position space by tailored electric and magnetic fields before being extracted into vacuum and re-accelerated~\cite{Taqqu:2006mv, Antognini:2021muCool}.

In recent years, this approach has gained renewed interest as a route to intense slow-muon beams for precision experiments. At PSI, the continuous surface-muon beams have a kinetic energy of approximately $4~\mathrm{MeV}$ and a transverse size of order $10~\mathrm{mm}$, whereas the muCool programme aims to deliver a sub-mm, transversely tagged beam of eV-energy muons that is subsequently re-accelerated by pulsed electric fields to the keV--MeV range. The targeted overall phase-space compression is a factor of $10^{9}$--$10^{10}$ at an efficiency of $2\times10^{-5}$--$2\times10^{-4}$ (limited mainly by muon decay during the ${\lesssim}\,8~\mu\mathrm{s}$ cooling cycle and by the stopping fraction), corresponding to a net gain in beam brightness of order $10^{5}$~\cite{Antognini:2021muCool,Antognini:2024xmb}. Applied to the $\pi$E5 beamline ($2.1\times10^{8}~\mu^{+}/\mathrm{s}$), this would yield ${\sim}2\times10^{4}~\mu^{+}/\mathrm{s}$ within an emittance of about $40~\mathrm{mm\,mrad}$ at $10~\mathrm{keV}$, and up to ${\sim}3\times10^{5}~\mu^{+}/\mathrm{s}$ at the planned High-Intensity Muon Beam (HIMB) upgrade~\cite{Aiba:2021bxe,Antognini:2021muCool}. Such a source would benefit a range of applications, including future muon Electric Dipole Moment (EDM) searches~\cite{Adelmann:2025nev}, muonium spectroscopy and gravity experiments~\cite{Mu-MASS:2021uou}, and high-resolution low-energy $\mu$SR.

\begin{figure}[htbp]
    \centering
    \includegraphics[width=0.9\linewidth]{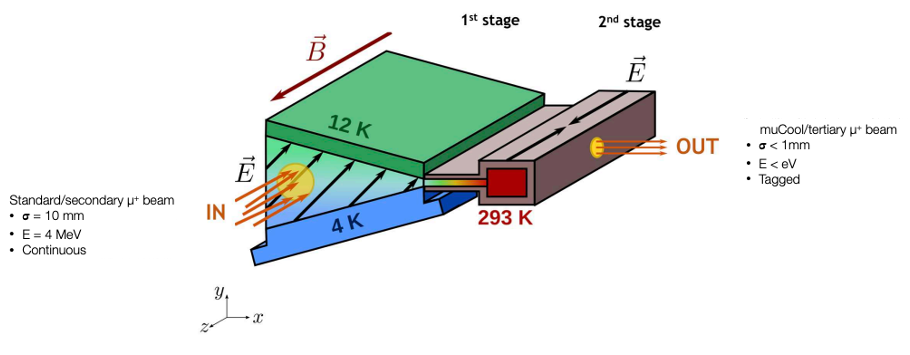}
    \caption{Principle of the muCool frictional-cooling scheme. A continuous surface-muon beam is stopped in a cryogenic, low-pressure helium target inside a 5-T solenoid. In the transverse (first) stage, a vertical gas-density gradient in crossed $\vec{E}\times\vec{B}$ fields compresses the swarm vertically while drifting it towards the second stage; there, an electric field with a longitudinal component compresses the swarm along the magnetic-field axis and guides it to a small orifice. The eV-energy muons are then extracted into vacuum and re-accelerated by pulsed electric fields to keV energies, yielding a tagged, sub-mm, high-brightness $\mu^{+}$ beam. Adapted from Ref.~\cite{Belosevic:2019pur}, with labels modified. \textcopyright{} The Author(s) 2019. Licensed under \href{https://creativecommons.org/licenses/by/4.0/}{CC BY 4.0}.}
    \label{fig:friction}
\end{figure}

The basic idea, first laid out in Ref.~\cite{Taqqu:2006mv} and illustrated in Figure~\ref{fig:friction}, is to stop the incoming beam in a few mbar of cryogenic helium gas inside a 5-T solenoid. At the resulting eV-scale energies, the muon motion is governed by the interplay between the cyclotron motion, the applied electric field, and the $\mu^{+}$--He collision frequency $\nu$: the drift velocity $\vec{v}_{D} \propto \frac{\mu E}{1+\omega^{2}/\nu^{2}}\left[\hat{E} + \frac{\omega}{\nu}\,\hat{E}\times\hat{B} + \frac{\omega^{2}}{\nu^{2}}\,(\hat{E}\cdot\hat{B})\,\hat{B}\right]$ depends on the local gas density through $\nu$. A vertical temperature gradient (from ${\sim}4~\mathrm{K}$ to ${\sim}12~\mathrm{K}$) therefore translates into a position-dependent drift direction, which, combined with a suitably shaped electric field, steers the entire swarm towards a single sub-mm spot at the target orifice. In this sense the approach achieves true phase-space compression through dissipation, rather than simple deceleration.

The muCool program has progressed through a staged demonstration of all key ingredients. The first milestone was the demonstration of longitudinal compression along the magnetic-field axis~\cite{Bao:2014xxa}: a $10~\mathrm{MeV}$/c $\mu^{+}$ beam was stopped in a few mbar of helium in a 5-T field, and the stopped swarm, initially extending over $16~\mathrm{cm}$, was compressed to a few mm within microseconds using a V-shaped electrostatic potential. Subsequent measurements at room temperature extended this to a $20~\mathrm{cm}$ stop distribution compressed to sub-mm extent within $2~\mu\mathrm{s}$, and additionally demonstrated the $\vec{E}\times\vec{B}$ drift of the compressed swarm perpendicular to the magnetic field, as required for transport towards the extraction point~\cite{Belosevic:2019pur}. The static gas-density gradient needed for the transverse stage was demonstrated separately by neutron radiography of a cryogenic $^{3}$He cell, achieving density ratios above 3~\cite{Wichmann:2016NIMA}.

Transverse compression was then demonstrated in a dedicated cryogenic target~\cite{Antognini:2020uyp}: the vertical density gradient produced a position-dependent drift velocity in crossed electric and magnetic fields, reducing a $12.5~\mathrm{MeV}$/c muon stop distribution extending over $14~\mathrm{mm}$ to $0.25~\mathrm{mm}$ (RMS) within $3.5~\mu\mathrm{s}$. Both the longitudinal and transverse measurements agree well with \textsc{Geant4}-based simulations that include the relevant low-energy $\mu^{+}$--He elastic and charge-exchange ($\mu^{+}\!\leftrightarrow\!$~muonium) cross sections, validating the microscopic transport model underlying the scheme.

Most recently, the collaboration reported the first demonstration of \emph{simultaneous} compression in both spatial dimensions (``mixed compression''), combining the transverse and longitudinal techniques within a single cryogenic target featuring an electric field with both transverse and longitudinal components~\cite{Antognini:2024xmb}. The evolution of the muon swarm was monitored with an array of positron detectors, and the measured time spectra show the muons drifting towards the target tip while being compressed in both the $y$ and $z$ directions. The validated simulations predict a compression efficiency of about $90\%$ (excluding muon decay) within $5~\mu\mathrm{s}$ for this stage. Together with earlier status reports~\cite{Lospalluto:2024M4F}, these results establish frictional cooling as an experimentally validated beam-manipulation technique rather than a purely conceptual proposal.

The remaining major step is the extraction of the compressed muons into vacuum through a ${\sim}1~\mathrm{mm}$ orifice, followed by pulsed re-acceleration and transport out of the solenoid. This stage is technically demanding because the extraction efficiency depends sensitively on the gas density profile, the gas flow through the orifice, and the electric-field configuration in the orifice region, all of which must preserve the small phase-space volume achieved during compression while minimizing losses from scattering and muonium formation. The operating conditions for this stage have been defined on the basis of the validated simulations, and a target incorporating the extraction and re-acceleration stages is being commissioned for beam tests at PSI~\cite{Antognini:2024xmb,Lospalluto:2024M4F}. Successful extraction would complete the transition from proof-of-principle demonstrations to a practical ultra-low-energy, high-brightness $\mu^{+}$ source.

The development of muCool illustrates the evolution of frictional cooling from a theoretical concept for phase-space reduction in matter into a staged experimental programme with well-defined milestones. With longitudinal, transverse, and now simultaneous two-dimensional compression experimentally demonstrated, the extraction and re-acceleration of the cooled muons remain the final steps towards a practical source for precision muon experiments, with its ultimate reach set by the intensity of the driving beamline and the planned HIMB upgrade at PSI~\cite{Aiba:2021bxe}.

\section{Thermal Muon Production via Laser Ionisation of Muonium}
\label{sec:laser}

Rather than cooling an already formed muon beam through repeated interactions with matter, laser-ionisation cooling produces an intrinsically low-emittance positive-muon beam through the formation and subsequent laser ionisation of thermal muonium ($\mu^{+}e^{-}$). Incident positive muons are stopped and thermalised in a suitable material, where they capture electrons to form muonium atoms. Muonium emitted from the material into vacuum is then resonantly ionised by lasers, releasing positive muons with a kinetic-energy distribution determined primarily by the thermal motion of the muonium atoms. The underlying concept was first proposed by Nagamine and Mills in 1986~\cite{Mills:1986zzb}, who observed the formation of thermal muonium by implanting positive muons into hot tungsten targets. This pioneering observation established the basis for producing low-energy muons through the formation and subsequent manipulation of muonium atoms.

\begin{figure}[htbp]
    \centering
    \includegraphics[width=0.68\linewidth]{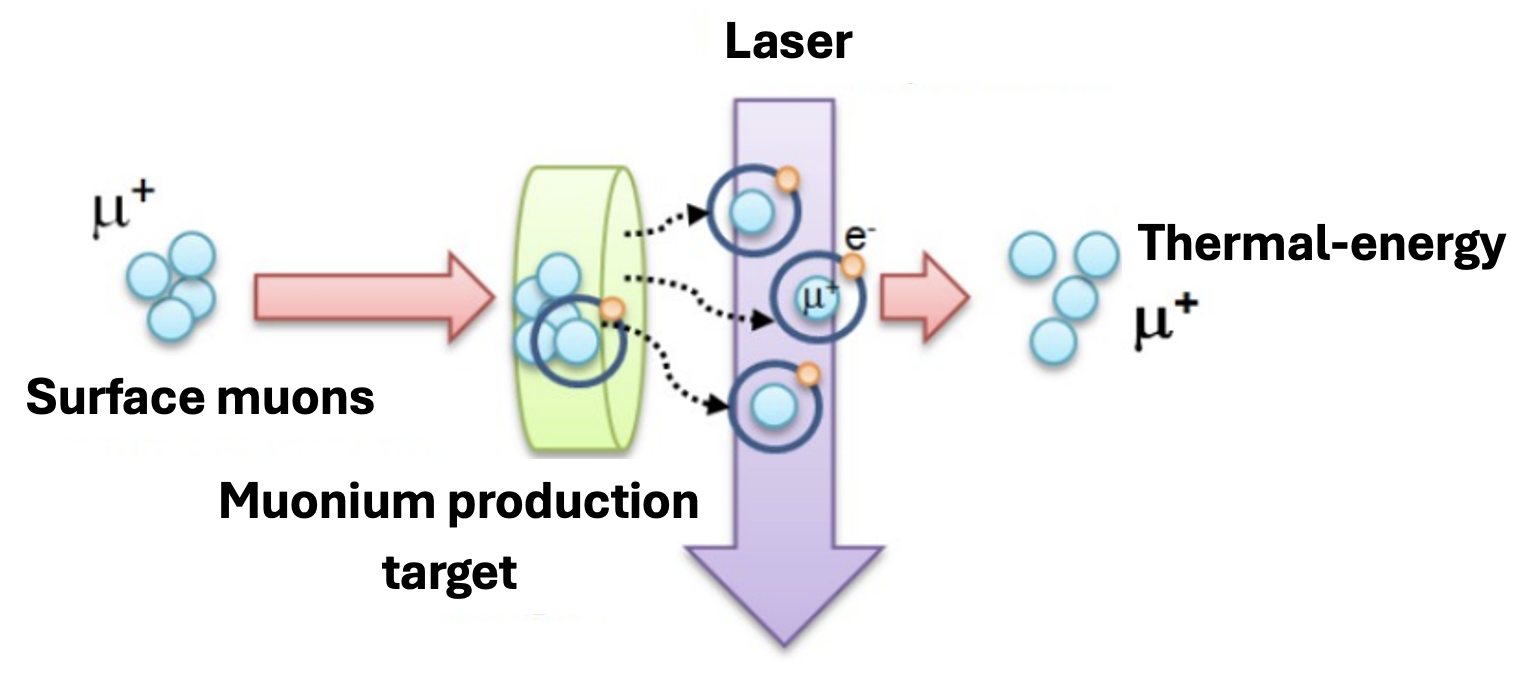}
    \includegraphics[width=0.31\linewidth]{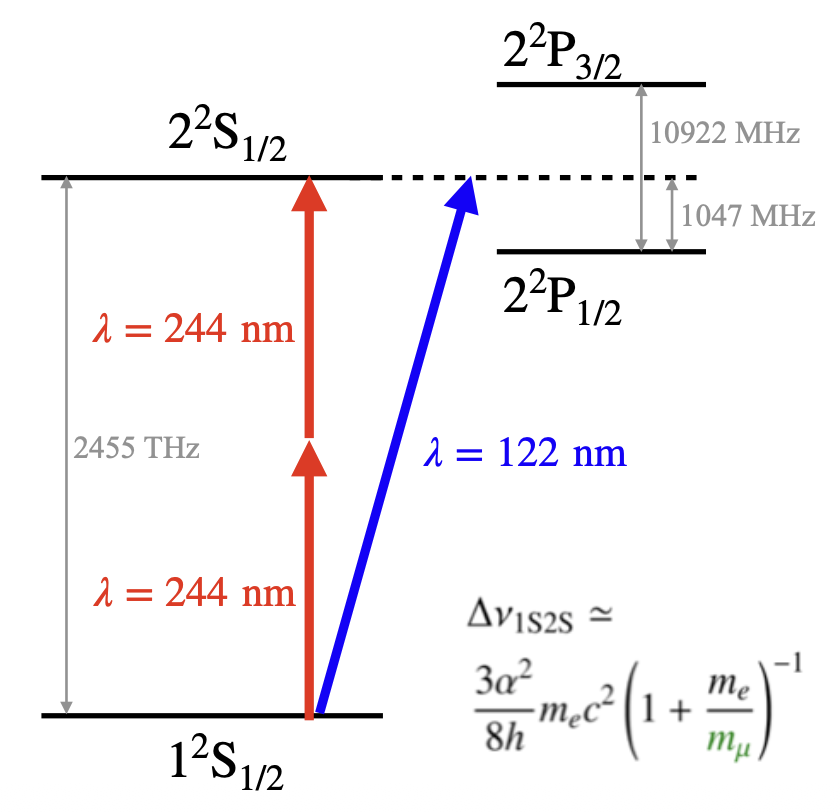}
    \caption{Left: Schematic principle of thermal muon production through the laser ionisation of thermal muonium. Surface $\mu^{+}$ are stopped and thermalised in a muonium-production target, where they capture electrons to form muonium atoms. Muonium emitted into vacuum is subsequently resonantly excited and photoionised, producing ultra-cold $\mu^{+}$ for subsequent acceleration. Right: Two alternative resonant-excitation schemes: single-photon excitation of the $1S$--$2P$ transition at 122~nm and two-photon excitation of the $1S$--$2S$ transition at 244~nm.}
    \label{fig:laser}
\end{figure}

The concept was further developed by the J-PARC Muon $g\!-\!2$/EDM collaboration~\cite{abe_new_2019}, for which the extremely small transverse emittance required to inject and store muons in a compact storage ring without electric-field focusing motivated the development of an ultra-low-emittance muon source. In this approach, thermal muons are generated through laser ionisation of muonium and subsequently re-accelerated while preserving their small emittance, yielding a beam suitable for precision measurements.
The production scheme, illustrated schematically in Fig.~\ref{fig:laser}, begins with conventional surface muons ($28~\mathrm{MeV}$/c) produced directly from pion decay near the surface of a production target~\cite{Otani:2015lra}. These positive muons are transported to a silica aerogel target, where they are stopped and capture electrons to form thermal muonium. A fraction of the muonium atoms diffuse through the aerogel and are emitted into the downstream vacuum, where they can subsequently be ionised by laser excitation.

The replacement of conventional hot-tungsten targets with silica-based aerogel represented a major breakthrough in the development of this technique. Although muonium production had previously been demonstrated using hot tungsten and silica powder~\cite{Marshall:1978sr,Janissen:1989gv}, hot tungsten exhibited a very low conversion efficiency, whereas the particulate nature of silica powder resulted in poor vacuum compatibility, limiting the practical applicability of both approaches. Considerable effort was therefore devoted to identifying alternative materials capable of providing a higher yield of muonium emission into vacuum. The first observation of muonium emission from silica aerogel was achieved at the TRIUMF muon beam facility~\cite{Bakule:2013poa}. A further major improvement was subsequently obtained through the development of laser-ablated silica-aerogel targets, whose tailored microstructure significantly enhanced muonium diffusion and emission into vacuum compared with untreated aerogel~\cite{Beer:2014ooa}. More recently, the J-PARC Muon $g\!-\!2$/EDM collaboration further optimised the laser-ablation process by applying it to both surfaces of the aerogel. This double-sided treatment mitigates the bending observed in the single-sided laser-ablated samples at TRIUMF, thereby improving the mechanical stability of the target. The developments at TRIUMF, together with the subsequent optimisation at J-PARC, have established the technological basis for the muonium production required by the J-PARC Muon $g\!-\!2$/EDM experiment.

The emitted muonium atoms are then resonantly excited using high-power laser systems and subsequently ionised to produce free positive muons with kinetic energies of around \(25~\mathrm{meV}\). Two laser excitation schemes have been explored: a Lyman-$\alpha$ transition at \(122~\mathrm{nm}\), corresponding to the muonium \(1S\)--\(2P\) transition~\cite{Nagamine:1995zz,Bakule:2008zz}, followed by a \(355~\mathrm{nm}\) light source for ionisation from the \(2P\) state; and a two-photon excitation scheme at \(244~\mathrm{nm}\) via the metastable \(1S\)--\(2S\) transition~\cite{Chu:1988zz}.
The \(122~\mathrm{nm}\) scheme generally provides a higher excitation efficiency, but the generation of high-power vacuum ultraviolet (VUV) light remains technically challenging. An excitation efficiency exceeding \(70\%\) has been adopted as the design target for the J-PARC Muon \(g\!-\!2\)/EDM experiment. Achieving this target requires a laser pulse energy of approximately \(100~\mu\mathrm{J}\) and a spectral linewidth of about \(80~\mathrm{GHz}\)~\cite{Saito:2016jtj,abe_new_2019}. In contrast, the \(244~\mathrm{nm}\) two-photon excitation scheme relies on higher laser intensity and therefore requires substantially larger pulse energies. An estimated value of around \(60~\mathrm{mJ}\) and the use of multiple mirror reflections can enhance the effective interaction intensity, making this scheme more accessible with existing laser technologies~\cite{Kamioka:2023xob}. It is therefore particularly well suited to an early proof-of-principle experiment aimed at demonstrating laser ionisation and validating the underlying production scheme. More recently, a major milestone was achieved at J-PARC, where thermal positive muons produced through two-photon excitation and subsequent ionisation of thermal muonium were successfully re-accelerated to \(100~\mathrm{keV}\) using a radio-frequency quadrupole cavity~\cite{aritome_acceleration_2025}. Details of muon acceleration at J-PARC are reviewed in a companion article in this issue~\cite{kamioka}.

The overall efficiency of the laser ionisation cooling approach is determined by several sequential processes~\cite{Zhang:2021cba}, including muonium formation (with a formation probability of 0.52 according to ~\cite{Beer:2014ooa}), diffusion and emission from the aerogel target, laser ionisation efficiency, and beam capture and acceleration. The diffusion process is modelled through a three-dimensional random walk of muonium atoms inside the aerogel, with the diffusion coefficient obtained from experimental data fitting. The overall efficiency for producing thermal muonium atoms within the laser irradiation region is estimated to be 0.0034~\cite{abe_new_2019}. During the muonium formation process, the initial muon polarization is reduced to approximately 50\%. Efforts have also focused on improving the efficiency and beam quality at individual stages of the production process, including the development of multilayer aerogel targets to enhance the muonium yield in vacuum and the laser pumping of muonium prior to ionisation to produce thermal muons with higher polarisation~\cite{Zhang:2022ilj,Venanzoni:2025rsd}.

This technology is also envisaged as a key component of the proposed $\mu$TRISTAN project~\cite{Hamada:2022mua}. By employing a high-power proton accelerator together with a pion-production storage ring that enables multiple target traversals, $\mu$TRISTAN aims to significantly enhance pion and surface-muon production rates. Combined with further improvements in thermal muon generation efficiency, laser ionisation cooling provides a promising route towards high-brightness muon beams for next-generation precision experiments and, potentially, future muon collider applications.

\section{Production of Slow Negative Muons}
\label{sec:negative}

The production of high-brightness low-energy negative muon beams presents a
significantly greater challenge than for positive muons. Although many beam
manipulation techniques have been successfully developed for $\mu^+$ beams,
most of them cannot be directly extended to $\mu^-$ because of the
fundamentally different interactions of negative muons with matter. Slow or stopped negative muons are rapidly captured by surrounding atoms
to form muonic atoms. Following the atomic cascade, the muon is eventually
captured by the nucleus through the weak interaction or decays while bound to
the atom. Consequently, unlike positive muons, there exists no neutral
intermediate state analogous to muonium that can escape from the material and
be subsequently ionised by lasers. This irreversible atomic capture makes the
production of ultra-slow $\mu^-$ beams substantially more difficult than that
of $\mu^+$ beams. As a result, alternative approaches based on direct
extraction or indirect cooling mechanisms have been explored over the past
several decades.

\subsection{Cyclotron-Trap Extraction}

One of the early approaches developed for producing slow negative muons employs a cyclotron trap to decelerate and extract low-energy $\mu^-$ beams. In this scheme, incoming muons with momenta of several tens of MeV/$c$ are confined in a magnetic trap and repeatedly pass through thin degrader foils, where they gradually lose kinetic energy. After sufficient deceleration, the muons are extracted from the trap with kinetic energies typically in the range of 10--50~keV. This technique enables the efficient pulsed production of low-energy negative muons. Although it does not provide significant transverse emittance reduction, cyclotron trapping offers a practical method for generating moderated $\mu^-$ beams suitable for precision muonic atom spectroscopy and other applications~\cite{Prokscha:2008zz}.

A representative implementation of this technique is the Low-Energy Negative Muon Channel at PSI, where the Muon Extraction Channel (MEC) was developed to provide pulsed low-energy $\mu^-$ beams for precision muonic atom spectroscopy. The extracted muons, with energies in the tens of keV range, were transported to experimental targets after moderation in the cyclotron trap. This facility played a crucial role in the measurement of the 2S Lamb shift in muonic hydrogen, leading to a precise determination of the proton charge radius~\cite{Antognini:2005fe}.

\subsection{Muon-Catalyzed Fusion as an Indirect Cooling Method}

A conceptually different approach exploits muon-catalyzed fusion ($\mu$CF)~\cite{Kamimura:2021msf} as an indirect mechanism for producing slow negative muons. Rather than slowing the incident muon through conventional moderation, this method takes advantage of the fact that the muon can be liberated following a fusion cycle.

After an incident $\mu^-$ is captured by a deuterium or tritium atom, muonic atoms are formed, which subsequently participate in the $\mu$CF cycle.
Following fusion, the muon is released with a kinetic energy of approximately 10~keV. If the fusion takes place sufficiently close to the surface of a thin hydrogen isotope layer, the liberated muon can escape into vacuum and be collected for subsequent transport. In this sense, the fusion process acts as an indirect beam cooling or moderation mechanism by regenerating free low-energy muons from initially energetic incident particles. 

This concept was originally proposed by Nagamine and
collaborators in the late 1980s~\cite{nagamine1989ultra}. Recent theoretical and experimental studies~\cite{Yamashita:2024inr,toyama2026direct} have revisited the expected energy spectrum and emission probability of muons released from thin deuterium--tritium films. Numerical calculations indicate that the emitted muons are concentrated around several keV to approximately 10~keV, providing an attractive starting point for subsequent beam cooling and focusing.

We note that the $\mu$CF scheme alone does not produce a beam with sufficiently small energy spread or transverse emittance for most applications. Consequently, development efforts~\cite{muhlbauer1996frictional} have focused on combining $\mu$CF with frictional cooling: negative muons released from the fusion target are first extracted by electric fields before entering a frictional cooling stage, where repeated energy loss in thin foils together with re-acceleration reduces the energy spread and improves the beam phase-space distribution. Recent conceptual designs further propose curved thin-film coolers, which simultaneously decelerate, focus and monochromatize the beam, offering significant improvements over conventional planar cooling structures~\cite{kamimura2023comprehensive}.

\section{Summary}

Muon beam cooling and moderation techniques are essential for expanding the scientific reach of muon-based experiments by improving beam brightness, reducing emittance, and providing access to low-energy muon beams. The optimal approach depends on the requirements of the application, ranging from high-brightness muon sources for future colliders to ultra-low-energy beams for precision spectroscopy and condensed-matter studies.

Among the various approaches, ionisation cooling remains the most mature
technique for high-energy muon beams and is the baseline technology for future muon colliders. In contrast, low-energy moderation and frictional cooling approaches target the production of slow muon beams. Laser ionisation of thermal muonium provides a complementary route by producing intrinsically low-emittance muon beams through atomic processes, and recent demonstrations of extraction and acceleration represent a significant milestone towards next-generation precision muon experiments.

The production of slow negative muons remains considerably more challenging because of nuclear capture. Emerging concepts based on muon-catalyzed fusion and subsequent cooling stages offer a potential pathway towards higher-brightness negative muon sources. Continued progress in high-intensity muon production, advanced beam manipulation techniques, and improved extraction and transport systems will be essential for realizing the full potential of future muon facilities.


\acknowledgments

The authors thank Patrick Strasser for helpful comments on negative-muon cooling and moderation. 
C.~Z. is supported by the Leverhulme Trust under Grant LIP-2021-014.
K.~S.~K. is supported by the Shanghai Pilot Program for Basic Research (Grant No. 21TQ1400221).
This document is endorsed by the IMCC. Some of the work described was funded by the European Union (EU). Views and opinions expressed are however those of the author(s) only and do not necessarily reflect those of the EU or European Research Executive Agency (REA). Neither the EU nor the REA can be held responsible for them.



\bibliographystyle{JHEP}
\bibliography{biblio}

\end{document}